\documentclass[a4paper,11pt]{article}
\usepackage{jinstpub} % for details on the use of the package, please see the JINST-author-manual
\usepackage{lineno}
\usepackage[T1]{fontenc}
\usepackage{siunitx}
\usepackage{physics}
\usepackage{graphicx}
\usepackage{xspace}
\usepackage{subcaption}
\usepackage{booktabs}
\usepackage{multirow}
\usepackage{hyperref}
\usepackage{orcidlink}
\usepackage{enumitem}
\usepackage{microtype}
\usepackage{xcolor}

\newcommand{\TempLow}{\SI{-33}{\celsius}\xspace}

\newcommand{\RuntimeRT}{\textasciitilde{}110\,days}

\title{Validation of the COSINE-100U NaI(Tl) Encapsulation for Low-Temperature Operation in Liquid Scintillator}

\author[1]{K. Park$^{*}$,}
\author[1]{S.-J. Cho,}
\author[2]{L. E. Fran{\c c}a}
\author[3]{C. Ha,}
\author[3,1]{J. Y. Kim,}
\author[1]{K. W. Kim,}
\author[1]{S. H. Kim,}
\author[4,1]{W. K. Kim,}
\author[5]{Y. J. Ko,}
\author[6,1]{D. H. Lee,}
\author[1,4]{H. S. Lee,}
\author[1]{I. S. Lee,}
\author[4,1]{S. H. Lee,}
\author[6,1]{S. D. Park,}
\author[1]{G. H. Yu}

\affiliation[1]{Center for Underground Physics (CUP), Institute for Basic Science (IBS), Daejeon 34126, Republic of Korea}
\affiliation[2]{Physics Institute, University of S\~{a}o Paulo, S\~{a}o Paulo 05508-090, Brazil}
\affiliation[3]{Department of Physics, Chung-Ang University, Seoul 06974, Republic of Korea}
\affiliation[4]{IBS School, University of Science and Technology (UST), Daejeon 34126, Republic of Korea}
\affiliation[5]{Department of Physics, Jeju National University, Jeju 63243, Republic of Korea}
\affiliation[6]{Department of Physics, Kyungpook National University, Daegu 41566, Republic of Korea}

\emailAdd{kihong@ibs.re.kr}
\emailAdd{hyunsulee@ibs.re.kr}
\abstract{
The COSINE-100U (upgrade) will enhance the sensitivity of the COSINE-100 dark matter search by operating the detector array immersed in liquid scintillator (LS) at $-30^{\circ}\text{C}$. To validate the detector design for these conditions, we constructed a module using the COSINE-100U encapsulation and performed a dedicated long-term stability study. The module was first monitored at room temperature for \RuntimeRT\ in air, followed by a one-week immersion in LAB-based LS to verify initial compatibility. Upon confirming stable optical performance, the temperature was lowered to $-33^{\circ}\text{C}$. During approximately 150 days of continuous operation at low temperature, we observed no degradation in performance. These results demonstrate the chemical and mechanical robustness of the encapsulation, confirming its suitability for the COSINE-100U physics run.
}


\begin{document}
\maketitle
\flushbottom

\section{Introduction}\label{sec:intro}
Astrophysical and cosmological observations consistently point to the presence of non-luminous matter that dominates the mass density of the Universe~\cite{ref:dm_review1,ref:dm_review2}. The nature of this dark matter remains unknown, and several complementary strategies are pursued in parallel: direct searches with low-background underground detectors, indirect probes of annihilation or decay products, and collider-based searches for dark-sector states at electron–positron and hadron colliders~\cite{ref:dd_review,ref:indirect_review,ref:collider_dm1,ref:collider_dm2,%
Park2023_SearchDP_JASS,Park2024_DarkPhotonML_JKPS}.

Although no concrete evidence of dark matter interactions has yet been observed, there remains the long-standing debate surrounding DAMA/LIBRA, which reported a clear annual-modulation signal in a NaI(Tl) crystal array compatible with dark matter~\cite{ref:dama1,ref:dama2}. The COSINE-100 experiment~\cite{ref:cosine_initial} was designed to test DAMA/LIBRA using the same target material and comparable analysis techniques. The experiment deployed eight low-background NaI(Tl) crystals, with a total mass of 106 kg, inside a liquid-scintillator veto~\cite{ref:cosine_ls} surrounded by copper, lead, and plastic scintillator muon panels~\cite{ref:cosine_muon} at the Yangyang underground laboratory in Korea. COSINE-100 has carried out both model-independent searches for a dark-matter–induced annual modulation~\cite{ref:cosine100_result2,ref:cosine100_result3,ref:cosine100_result6,ref:cosine100_result4} and model-dependent spectral analyses~\cite{ref:cosine100_result1,ref:cosine100_result7,ref:cosine100_result5}, finding no signals compatible with DAMA/LIBRA in NaI(Tl).

Beyond testing the DAMA/LIBRA claim, NaI(Tl) offers unique advantages due to the unpaired protons in both sodium and iodine isotopes. Given the relatively low mass number of sodium, NaI(Tl) has competitive sensitivity to Spin-Dependent WIMP-proton interactions in the low-mass dark matter region, as demonstrated in recent COSINE-100 results~\cite{ref:cosine100_result5}. To further explore this parameter space, the COSINE-100 upgrade (COSINE-100U)~\cite{ref:cosine100u_status} is being commissioned at Yemilab~\cite{ref:yemilab1,ref:yemilab2}, a new underground laboratory in Korea. This upgrade aims to enhance sensitivity to low-mass dark matter by increasing the light collection efficiency

A key feature of the upgrade is a novel crystal encapsulation technique~\cite{ref:neon_encap1}. Because NaI(Tl) is hygroscopic, crystals are typically encased in copper with a quartz optical window. To maximize light yield, we developed an encapsulation method that couples photomultiplier tubes (PMTs) directly to the crystal, eliminating the quartz window. This design enhances light collection efficiency by approximately 50\%~\cite{ref:neon_encap1,ref:neon_encap2}. The long-term stability of this design in liquid scintillator (LS) at room temperature (RT) was previously demonstrated by the NEON experiment~\cite{ref:neon_initial} over two years of operation~\cite{ref:neon_encap2}.

While COSINE-100U has begun initial physics operations at Yemilab~\cite{ref:yemilab1,ref:yemilab2} with crystals at RT, the ultimate goal is to operate the array at a low temperature (LT) of approximately $-30^{\circ}\text{C}$~\cite{ref:cosine100u_status}. This lower temperature offers the dual advantages of increased intrinsic light yield and improved pulse shape discrimination~\cite{LEE2022102709}. However, operating hygroscopic NaI(Tl) crystals immersed in LS at cryogenic temperatures introduces significant engineering challenges, specifically regarding the chemical compatibility of the encapsulation seals and the thermo-mechanical stress during cooling. 

This work serves as the critical validation of the encapsulation design prior to the cryogenic deployment of the full COSINE-100U array. We present a dedicated long-term stability measurement of a NaI(Tl) crystal encapsulated using the COSINE-100U method. The detector was operated first at RT without LS, and then fully immersed in LS at $-33^{\circ}\text{C}$. We monitored the light yield and energy resolution of the 59.54 keV $\gamma$-ray line from an $^{241}\text{Am}$ source throughout the study. We find that the optical quality of the NaI(Tl) crystal remained unchanged during an operation period of approximately one year, which included roughly 110 days at RT and 150 days at $-33^{\circ}\text{C}$. These results validate the robustness of the COSINE-100U encapsulation for operation in LS at approximately $-30^{\circ}\text{C}$.

\section{Experimental setup}\label{sec:setup}
The prototype detector utilizes the NaI-037 NaI(Tl) crystal~\cite{Lee2023_Frontiers_ULB_NaI}, grown at the Center for Underground Physics (IBS, Korea) using the purified powder described in Refs~\cite{ref:purification1,ref:purification2}. The crystal is cylindrical, with a diameter of 70\,mm and a height of 51\,mm, and has a mass of approximately 0.70\,kg after surface polishing.

We followed the encapsulation procedure developed for COSINE-100U~\cite{ref:cosine100u_status}, with one modification: the crystal edges were not machined, as the 70\,mm diameter already matches the active area of the 3-inch PMTs.
Prior to assembly, all encapsulation materials were cleaned in dilute Citranox via sonication, baked in an oven, and transferred to a glovebox. The glovebox atmosphere was maintained with a humidity (H$_2$O) level of $\mathcal{O}(10)$ ppm using a molecular-sieve trap. During the assembly, the glovebox was flushed with N$_2$ gas at a rate of 5 L/min to minimize $^{222}\text{Rn}$ contamination from component emanation. The crystal end faces were mirror-polished using a polishing pad with SiO$_2$ abrasives, while the barrel surface was gently wiped with anhydrous ethanol.

\begin{figure}[t]
  \centering
  \includegraphics[width=1.0\linewidth]{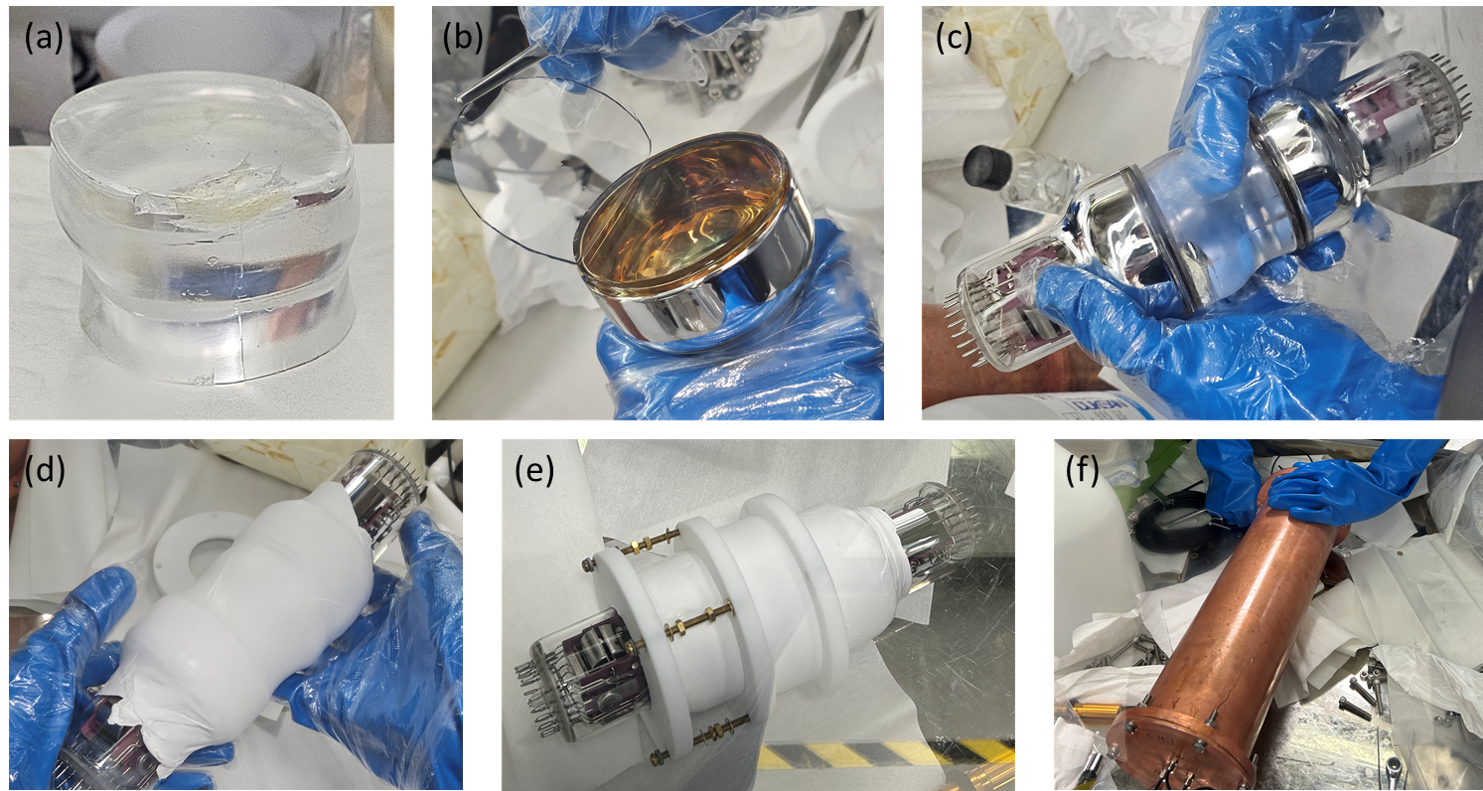}
  \caption{Assembly sequence of the NaI-037 NaI(Tl) detector: (a) the polished crystal after surface treatment; (b) attachment of the silicone optical pad to the PMT window; (c) the crystal coupled to two 3-inch Hamamatsu R12669SEL PMTs; (d) wrapping of the crystal and PMT assembly with PTFE-based Tetratex reflector; (e) installation of the PTFE holder to secure the crystal and PMTs; and (f) insertion of the assembled detector into the upgraded COSINE-100U copper housing.}
  \label{fig:crystal_assembly}
\end{figure}

\begin{figure}[t]
  \centering
  \includegraphics[width=0.9\linewidth]{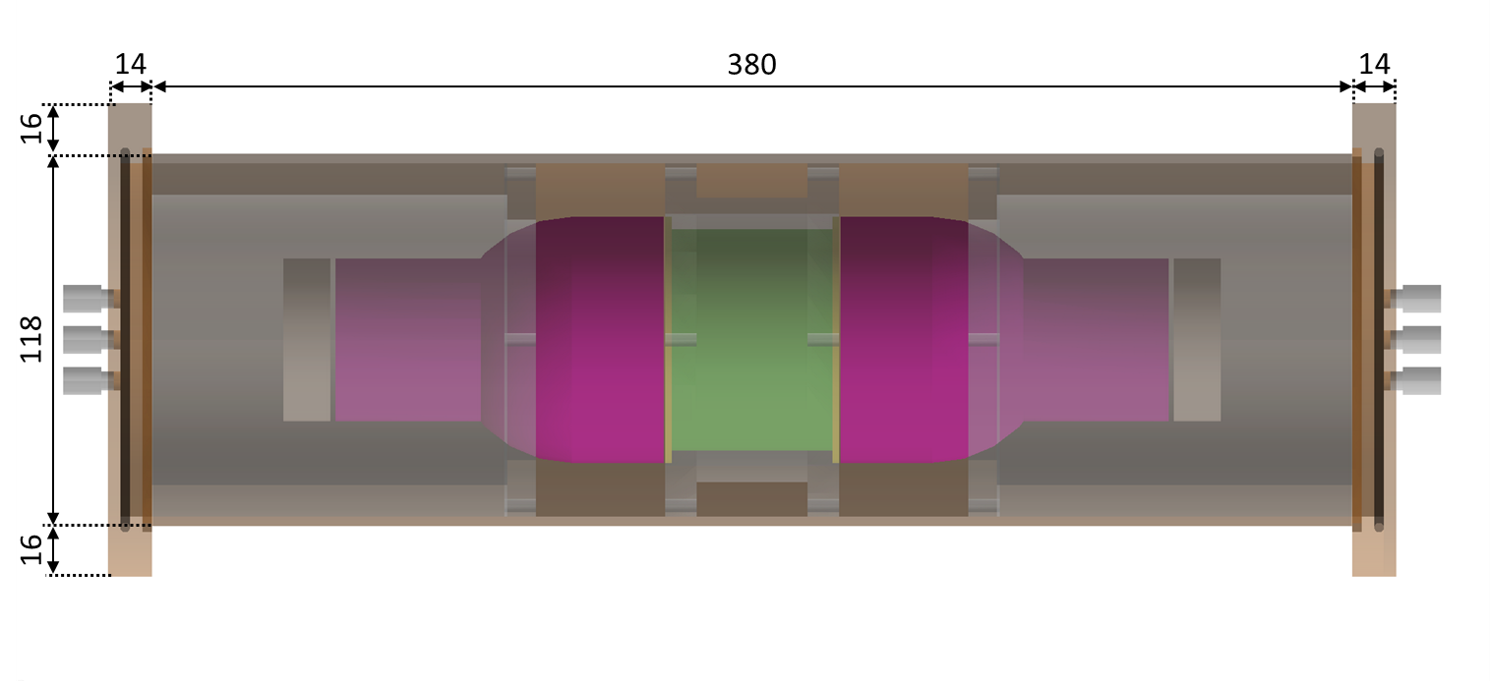}
  \caption{Schematic cross-sectional view of the fully encapsulated NaI-037 detector geometry. The NaI(Tl) crystal (green) is optically coupled to two 3-inch PMTs (magenta), with the PMT bases shown in gray. O-rings (black) at both lid--body interfaces provide the seal between the copper lids and the copper body. A PTFE (Teflon) support structure holds the crystal--PMT assembly inside the upgraded COSINE-100U copper housing; the main encapsulation dimensions are indicated in units of mm.}
    \label{fig:encap_schematic}
\end{figure}

For the optical interface, 2 mm thick silicone pads (Eljen EJ-560) were coupled to the top and bottom faces of the crystal. Two high-quantum-efficiency 3-inch PMTs (Hamamatsu R12669SEL) were attached to these pads to maximize light collection. The curved side of the crystal was wrapped in several layers of soft, Polytetrafluoroethylene (PTFE)-based Tetratex, which serves as a diffuse reflector to contain scintillation light. A PTFE holder secures the mechanical coupling between the crystal and PMTs, ensuring uniform optical contact. The entire assembly was centered in a copper support and enclosed in a copper housing matching the COSINE-100U design~\cite{ref:cosine100u_status}. Crucially, the housing is sealed using Viton O-rings selected for their chemical compatibility with LS and mechanical resilience at $-30^{\circ}\text{C}$. The overall assembly procedure is illustrated in Fig.~\ref{fig:crystal_assembly}.
Figure~\ref{fig:encap_schematic} shows a schematic of the completed copper encapsulation with the main dimensions indicated in units of mm. The copper housing is sealed using O-rings at both end-cap (lid--body) interfaces to ensure an airtight environment and prevent the ingress of ambient humidity.

\begin{figure}[t]
  \centering
  \includegraphics[width=0.9\linewidth]{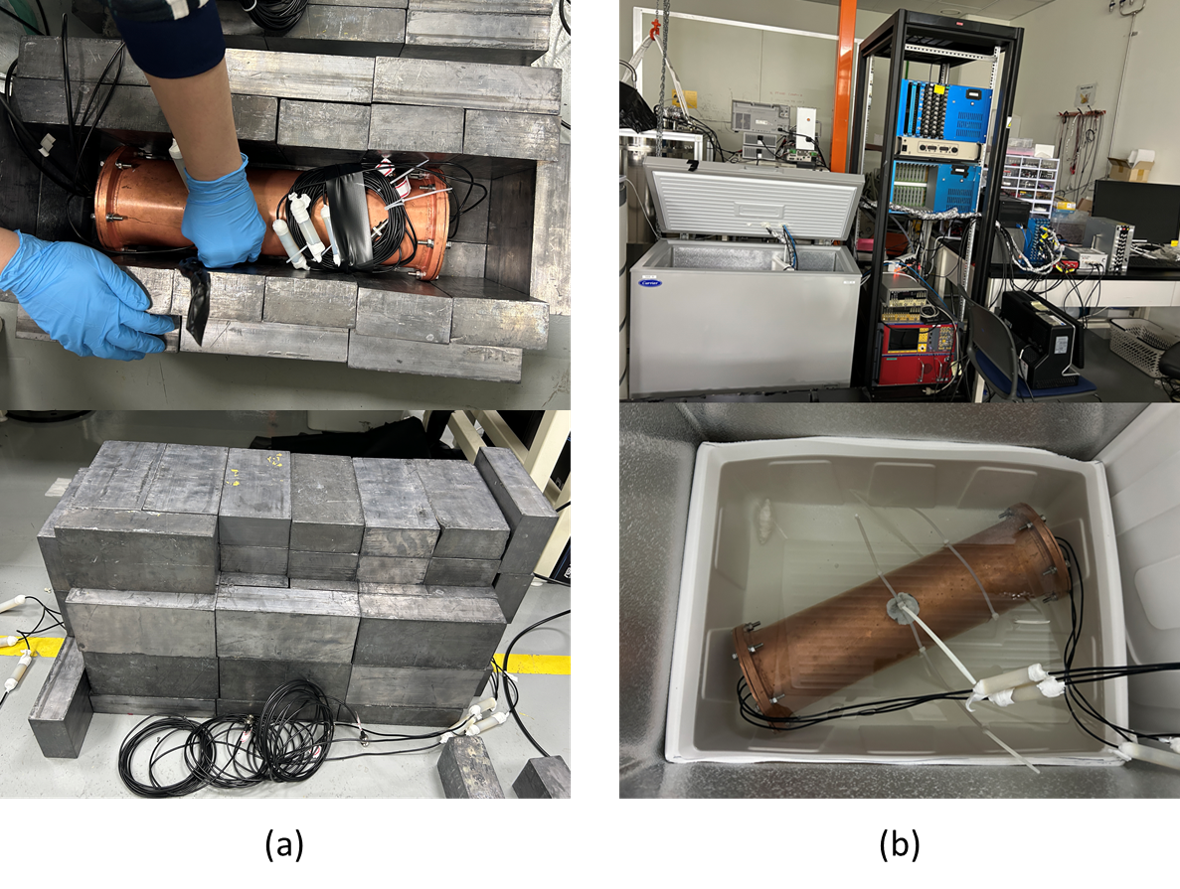}
  \caption{
  Photographs of the NaI-037 experimental setup. (a) Room temperature configuration: the encapsulated detector placed inside the lead shielding (top) and the lead bricks surrounding the setup (bottom). (b) Low-temperature configuration: the DAQ rack and refrigerator used for measurements (top) and the NaI-037 detector immersed in LS inside the refrigerator at $-33^{\circ}\text{C}$ (bottom).}
  \label{fig:setup_rt_lt}
\end{figure}

The stability of this encapsulation design in LS at RT was previously demonstrated by the NEON experiment over a two-year period~\cite{ref:neon_encap2}. Therefore, we first operated the detector in an ambient air environment at RT to verify the seal integrity against humidity. Since NaI(Tl) is hygroscopic, any ingress of ambient air would degrade the crystal surface and reduce the light yield. As shown in Fig.~\ref{fig:setup_rt_lt} (a), the detector was shielded by 10\,cm of lead to reduce external background radiation.

After approximately 110 days of stable operation in air, the detector was immersed in LS to prepare for low-temperature testing. Following a one-week stability check in LS at RT, the setup was transitioned to low-temperature operation. The detector and LS volume were placed in a plastic container inside a commercial chest refrigerator, as shown in Fig.~\ref{fig:setup_rt_lt} (b). To stress-test the encapsulation against thermal shock, the refrigerator was set directly to its minimum temperature without a gradual cooling ramp. The equilibrium temperature was measured at $-33^{\circ}\text{C}$, sufficient to validate the design for the planned $-30^{\circ}\text{C}$ operation of COSINE-100U. The detector was then operated under these conditions for approximately 150\,days to monitor long-term stability.

Signals from the PMTs were amplified by custom-built preamplifiers and digitized by a 500\,MHz, 12-bit flash analog-to-digital converter. A trigger was generated when a signal corresponding to one or more photoelectrons occurred in each PMT within a 200\,ns coincidence window. For RT data, an 8\,$\mu$s waveform (starting 2.4\,$\mu$s before the trigger) was recorded. For low-temperature operation, the recording window was extended to 16\,$\mu$s to account for the increased scintillation decay time of NaI(Tl) at $-33^{\circ}\text{C}$~\cite{LEE2022102709}. Data were transferred to a Linux computer via USB-3 and stored using a ROOT~\cite{ref:root}-based data acquisition software similar to the COSINE-100 DAQ system~\cite{COSINE-100:DAQ}.

The detector response was calibrated using 59.54\,keV $\gamma$-rays from an $^{241}\text{Am}$ source. To ensure the consistency of the long-term stability measurement, the source was maintained in a fixed position relative to the crystal throughout the entire campaign, eliminating systematic uncertainties related to position dependence.

\section{Data analysis and results}\label{sec:results}
\subsection{Scintillation characteristics at room temperature and low temperature}
\subsubsection{Decay time measurement}
The scintillation decay time directly influences the optimization of the integration window, the pile-up probability, and the efficiency of pulse-shape–based background rejection. To quantify these effects, we compared the scintillation decay times of 59.54 keV $\gamma$-ray events from an $^{241}\text{Am}$ source at RT and at $-33^{\circ}\text{C}$.

For a precise comparison, we constructed accumulated waveforms of the 59.54 keV peak events for both temperatures, as shown in Fig. 3. At RT, the waveform is well-described by a two-component exponential model~\cite{LEE2022102709}:
\begin{equation}
  V_{\mathrm{RT}}(t) =
    A_{1} e^{-(t - t_{0})/\tau_{1}}
  + A_{2} e^{-(t - t_{0})/\tau_{2}} \, ,
  \label{eq:two_exp_rt}
\end{equation}
where $\tau_1$ and $\tau_2$ denote the fast and slow decay constants, and $A_1$ and $A_2$ are the corresponding amplitudes.

\begin{figure}[t]
  \centering
  \includegraphics[width=0.9\linewidth]{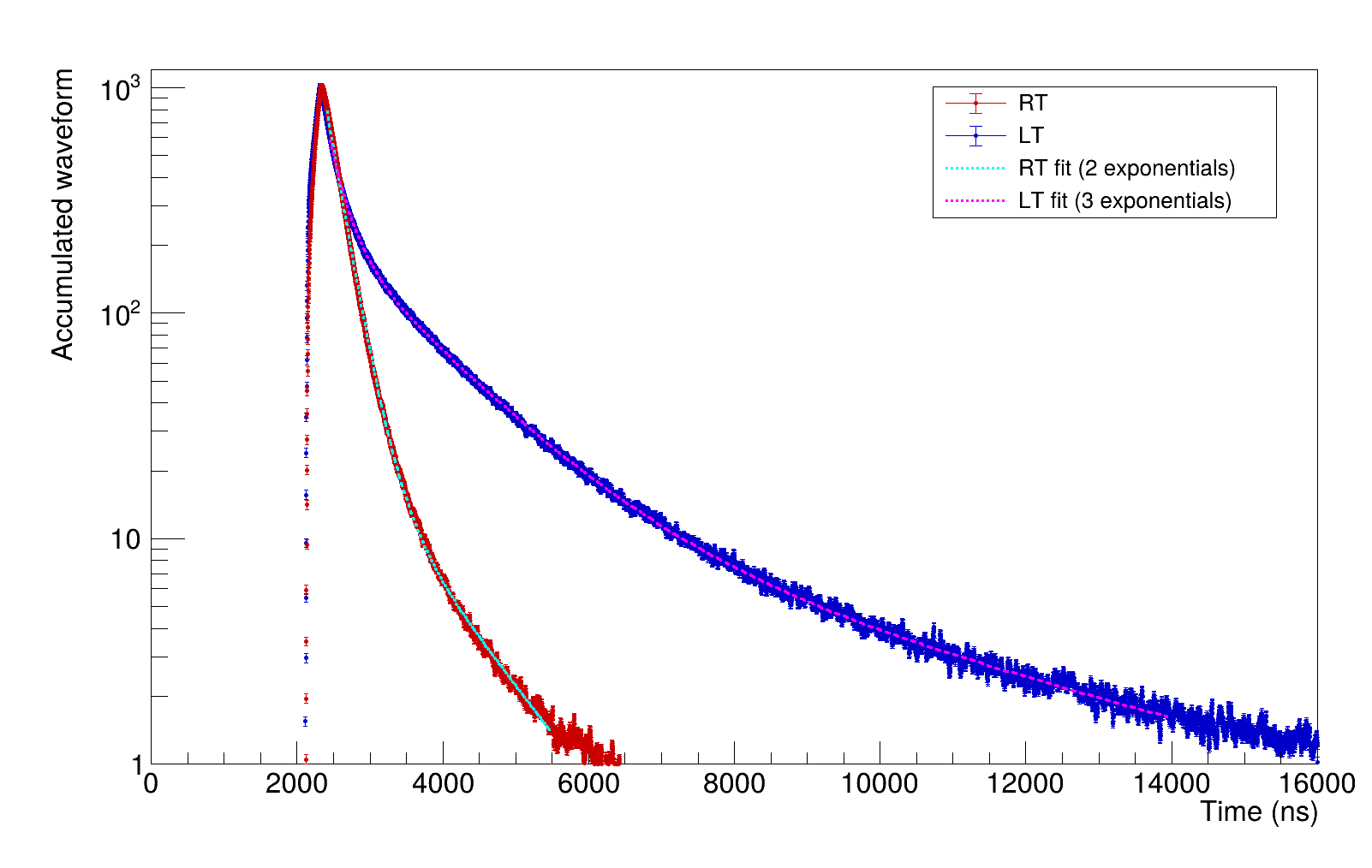}
  \caption{ Accumulated waveforms for 59.54 keV events from an $^{241}\text{Am}$ source at room temperature (RT) and $-33^{\circ}\text{C}$ (LT), with the corresponding best-fit multi-exponential models overlaid. The extended tail observed at $-33^{\circ}\text{C}$ reflects the increased contribution of the intermediate and slow decay components.}
  \label{fig:decay_waveform}
\end{figure}

At $-33^{\circ}\text{C}$, the same two-component model failed to adequately describe the waveform, particularly the extended decay tail. Consequently, we introduced a three-component model~\cite{ref:CsITl_LT}:
\begin{equation}
  V_{\mathrm{LT}}(t) =
    B_{1} e^{-(t - t_{0})/\tau_{1}}
  + B_{2} e^{-(t - t_{0})/\tau_{2}}
  + B_{3} e^{-(t - t_{0})/\tau_{3}} \, ,
  \label{eq:three_exp}
\end{equation}
where $\tau_1$, $\tau_2$, and $\tau_3$ represent the fast, intermediate, and slow decay constants, respectively, and $B_1$, $B_2$, and $B_3$ are the corresponding amplitudes. We fitted the accumulated data for both room-temperature and low-temperature configurations using these functions, as illustrated in Fig.~\ref{fig:decay_waveform}.

The extracted decay constants and component fractions are summarized in Table~\ref{tab:tau_summary}.
The component fractions are defined from the fitted exponential components using the relative areas of each term: $f_i = 100\times A_i\tau_i/(A_1\tau_1 + A_2\tau_2)$ for the RT two-exponential model in Eq.~(\ref{eq:two_exp_rt}), and $f_i = 100\times B_i\tau_i/(B_1\tau_1 + B_2\tau_2 + B_3\tau_3)$ for the LT three-exponential model in Eq.~(\ref{eq:three_exp}); therefore, $f_1+f_2=100\%$ (RT) and $f_1+f_2+f_3=100\%$ (LT). The fast decay constant ($\tau_1$) remains consistent at approximately 200 ns at both temperatures. The intermediate decay constant ($\tau_2$) is also stable, with a value on the order of 1000 ns. However, the low-temperature waveform exhibits an additional slow component ($\tau_3$) on the order of 5000 ns. Notably, the contribution of the fast component decreases dramatically from 89\% at RT to 29\% at $-33^{\circ}\text{C}$.

\begin{table}[t]
  \centering
  \small
  \setlength{\tabcolsep}{6pt}
  \caption{Fitted decay constants and component fractions for the multi-exponential model applied to accumulated
  $^{241}\text{Am}$ 59.54\,keV waveforms at room temperature and $-33^{\circ}\text{C}$.
  %The fractions are defined as $f_i = 100\times A_i/(A_1+A_2)$ for RT and $f_i = 100\times B_i/(B_1+B_2+B_3)$ for LT.
  At room temperature, a two-exponential model is used, so $f_3$ is not defined.
  Quoted uncertainties are statistical and rounded to one decimal place.}
  \label{tab:tau_summary}
  \begin{tabular}{lccccccc}
    \hline
    Temperature  & $\tau_{1}$ [ns] & $\tau_{2}$ [ns]  & $\tau_{3}$ [ns] & $f_{1}$ [\%] & $f_{2}$ [\%] & $f_{3}$ [\%] \\
    \hline
    Room temp. &  $220.4 \pm 0.6$ & $1030.0 \pm 4.3$ & $-$ & $88.8 \pm 0.1$ & $11.2 \pm 0.1$ & $-$ \\
    $-33^\circ$C & $208.9 \pm 2.4$ & $1244.1 \pm 4.7$ & $5048.8 \pm 29.6$ & $29.4 \pm 0.4$ & $54.5 \pm 0.2$ & $16.1 \pm 0.4$ \\
    \hline
  \end{tabular}
\end{table}

At lower temperatures, non-radiative quenching channels are suppressed, and a larger fraction of excitations are delayed in trapping states before reaching the Tl activator sites~\cite{LEE2022102709,ref:nai_temp2,ref:nai_temp3}. This results in a redistribution of scintillation light into the intermediate and slow components, leading to an overall longer decay time at $-33^{\circ}\text{C}$. This behavior necessitates a longer integration window for low-temperature operation and suggests that pulse-shape discrimination techniques may benefit from the enhanced separation between fast and slow components in the COSINE-100U environment.

\subsubsection{Light yields}

The single-photoelectron charge distribution is constructed from the charges of individual isolated clusters identified in the decay tail of the waveforms~\cite{Kims:2005dol}. These cluster charges are fitted with a model consisting of a Poisson signal component and an exponential background. The light yield of the detector is determined by comparing the mean charge of the 59.54\,keV peak to the mean single-photoelectron charge. The energy resolution is defined as the ratio $\sigma / \mu$, where $\sigma$ and $\mu$ correspond to the root-mean-square and mean, respectively, of the 59.54\,keV $\gamma$-ray charge distribution.

The measured light yield and energy resolution for the 59.54 keV line at RT and $-33^{\circ}\text{C}$ are summarized in Table~\ref{tab:LY_resolution_LS}. At RT, we measured a high light yield of $21.9 \pm 0.2$ photoelectrons/keV (PEs/keV). At LT, this value increased to $23.2 \pm 0.2$ PEs/keV. Correspondingly, the energy resolution improved from $4.02 \pm 0.15\%$ to $3.66 \pm 0.13\%$, a performance enhancement consistent with previous low-temperature studies~\cite{LEE2022102709}.

\begin{table}[t]
  \centering
  \caption{ Light yield and energy resolution for the $^{241}\text{Am}$ 59.54\,keV peak measured at room temperature and $-33^{\circ}\text{C}$
}
  \label{tab:LY_resolution_LS}
  \begin{tabular}{lccc}
    \hline
    Temperature & Window [$\mu$s] & Light yield [PEs/keV] & Resolution [\%] \\
    \hline
    Room temp.   & 8  & $21.9 \pm 0.2$ & $4.02 \pm 0.15$ \\
    \TempLow\  & 16 & $23.2 \pm 0.2$ & $3.66 \pm 0.13$ \\
    \hline
  \end{tabular}
\end{table}

\subsection{Stability of the detector encapsulation}
We evaluated the long-term stability of the NaI-037 light yield through frequent monitoring of the 59.54\,keV $\gamma$-ray peak. The monitoring campaign began in December 2024 with the detector operating in air at RT, utilizing the setup described in Fig.~\ref{fig:setup_rt_lt} (a). This phase continued until March 2025, after which the detector remained in ambient air conditions until June 2025. This initial period served to verify the mechanical air-tightness of the seal against ambient humidity.

In June 2025, the detector was immersed in LAB-based LS as shown in Fig.~\ref{fig:setup_rt_lt} (b). We first performed a one-week stability check in LS at RT to ensure no immediate chemical interaction or leakage occurred at the O-ring interface. Following this confirmation, the refrigerator was switched on and set directly to its minimum temperature to simulate a rapid cool-down scenario. We allowed approximately one week for the system to reach a stable thermal equilibrium at $-33^{\circ}\text{C}$. The low-temperature measurement phase spanned approximately 150 days, from July 2025 to December 2025, representing the critical validation period for the upgrade.

\begin{figure}[t]
  \centering
  \begin{subfigure}{\linewidth}
    \centering
    \includegraphics[width=\linewidth]{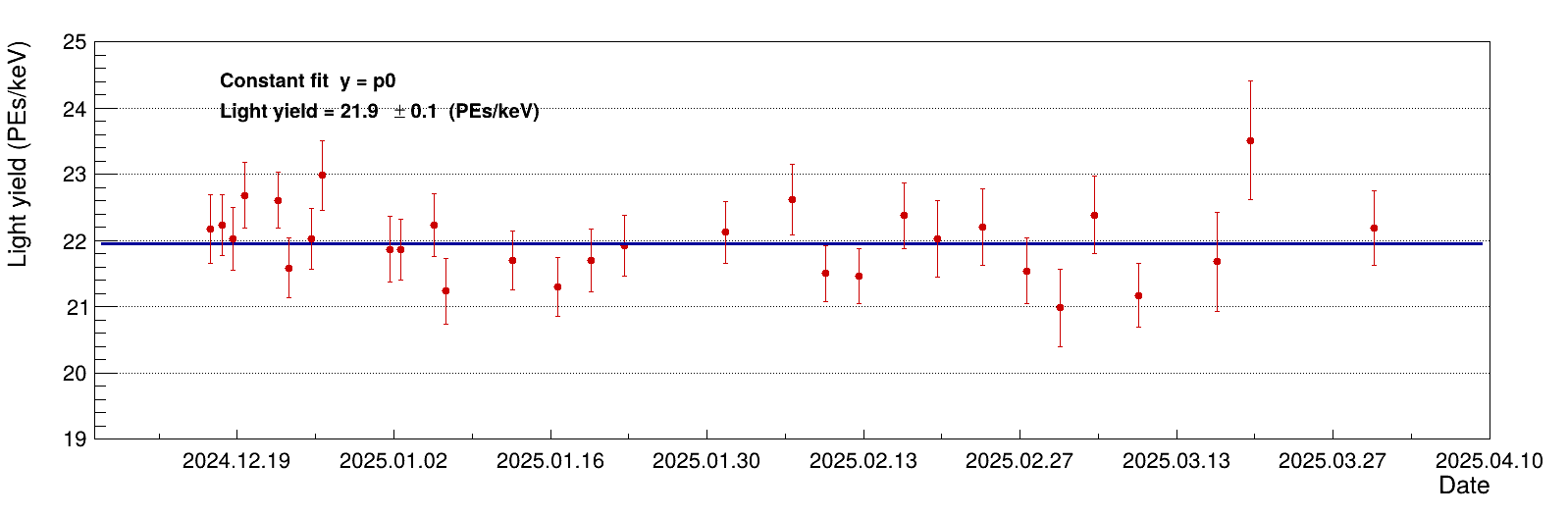}
    \caption{Room temperature}
    \label{fig:ly_stability_rt}
  \end{subfigure}
  \vspace{0.5em}
  \begin{subfigure}{\linewidth}
    \centering
    \includegraphics[width=\linewidth]{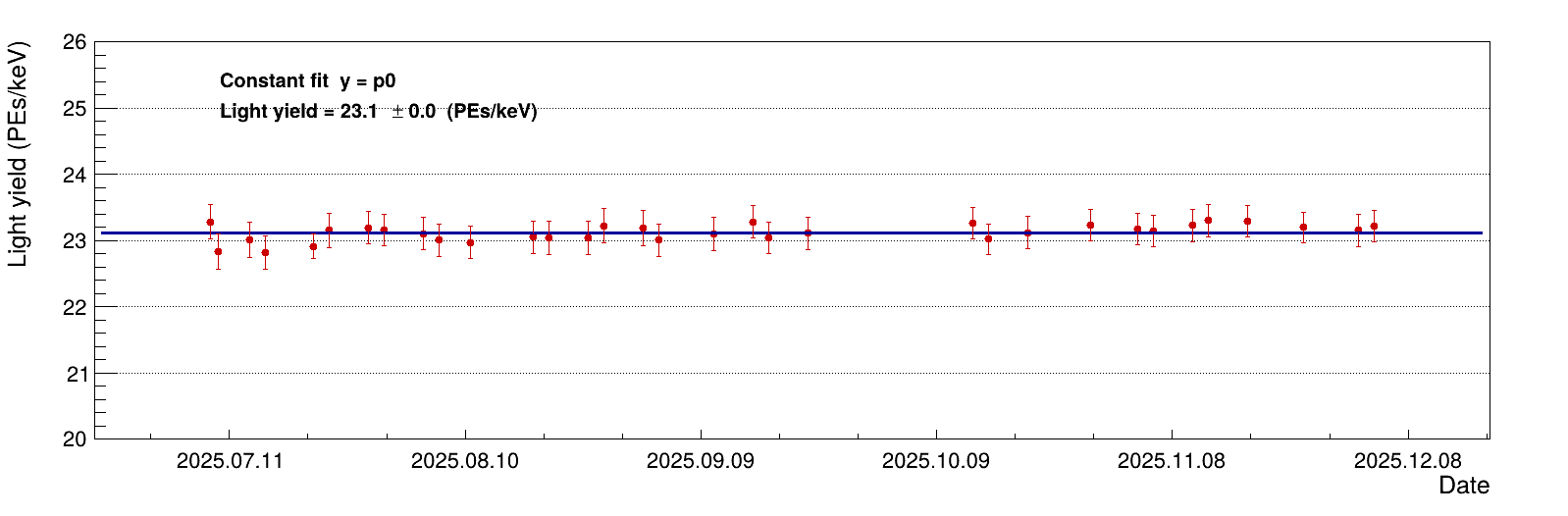}
    \caption{$-33^{\circ}\text{C}$}
    \label{fig:ly_stability_lt}
  \end{subfigure}

  \caption{ Light yield monitoring of the NaI-037 detector. (a) Time evolution of the light yield at room temperature in ambient air conditions. (b) Time evolution of the light yield at $-33^{\circ}\text{C}$ while immersed in LS. The solid lines represent fits to a constant function. The consistent light yield observed throughout the measurement periods demonstrates that the COSINE-100U encapsulation is sufficiently stable to maintain optical performance during long-term operation in LS at $-30^{\circ}\text{C}$. }
  \label{fig:ly_stability}
\end{figure}

Figure~\ref{fig:ly_stability} illustrates the temporal evolution of the measured light yield for the room-temperature (a) and $-33^{\circ}\text{C}$ (b) periods. The data points show excellent agreement with a constant light yield model. For the room-temperature phase, we measured a mean light yield of $21.94 \pm 0.09$ PEs/keV. For the crucial low-temperature phase in LS, the mean light yield was $23.11 \pm 0.04$ PEs/keV. The stability of these values confirms that the detector response is time-independent.

To quantitatively bound any potential degradation, we also fitted the data with a linear function ($y = p_0 + p_1 x$). The slope parameter $p_1$ provides an upper limit on the degradation rate. We obtained a value of $p_1 = (1.5 \pm 0.9)\times10^{-3}\ $ PEs/keV/day for $-33^{\circ}\text{C}$, which is consistent with zero within statistical uncertainties. This measurement effectively rules out common failure modes associated with LS immersion, such as the infiltration of LS into the PTFE reflector (which would reduce reflectivity) or the clouding of the crystal surface due to chemical ingress. Furthermore, the stability at $-33^{\circ}\text{C}$ confirms that the silicone optical pads maintained robust coupling to the crystal and PMTs despite the differential thermal contraction of the assembly.

Crucially, the stable operation in LS at $-33^{\circ}\text{C}$ for approximately 150 days provides strong evidence for the robustness of the COSINE-100U encapsulation. This result validates the engineering design for the full experiment's physics run at $-30^{\circ}\text{C}$, which is scheduled to commence in March 2026.

\section{Conclusions}\label{sec:conclusion}
We have successfully validated the performance and long-term stability of the COSINE-100U encapsulation design using the prototype NaI-037 detector. This study serves as a critical engineering pre-test for the experiment's upgrade, specifically targeting the challenges of operating hygroscopic NaI(Tl) crystals immersed in LS at cryogenic temperatures.

The detector exhibited robust performance across all testing phases. During the initial 100-day operation in air at RT, the light yield remained stable, confirming the mechanical integrity of the seals against ambient humidity. Following immersion in LS and a rapid cool-down to $-33^{\circ}\text{C}$, the detector maintained stable performance over a 150-day monitoring period. A linear fit to the light yield data during this phase is consistent with zero degradation, effectively ruling out chemical incompatibility with the LS or mechanical decoupling due to thermal stress.

In addition to demonstrating stability, the low-temperature operation provided clear performance benefits. At $-33^{\circ}\text{C}$, we observed a relative increase in light yield of approximately $5.6\%$ and an improvement in 59.54 keV energy resolution of approximately $9\%$ compared to RT. Analysis of the accumulated waveforms confirmed the expected increase in the scintillation decay time at LT, necessitating the extension of the integration window to 16 $\mu$s.

In summary, these results demonstrate that the upgraded encapsulation ensures chemical and mechanical robustness under the target operating conditions of COSINE-100U. With the stability of the design verified, the COSINE-100U experiment is on schedule to commence its physics run at $-30^{\circ}\text{C}$ in March 2026, with enhanced sensitivity to low-mass dark matter.

\acknowledgments
This work is supported by the Institute for Basic Science (IBS) under project code IBS-R016-A1,  NRF-2021R1A2C3010989, NRF-2021R1A2C1013761, RS-2024-00356960, RS-2025-25442707 and RS-2025-16064659, Republic of Korea; Grant No. 2022/13293-5 and 2025/01639-2 FAPESP, CNPq 304658/2023-5, Brazil.

% Bibliography

%% [A] Recommended: using JHEP.bst file
%% \bibliographystyle{JHEP}
%% \bibliography{biblio.bib}

%% or
%% [B] Manual formatting (see below)
%% (i) We suggest to always provide author, title and journal data or doi:
%% in short all the informations that clearly identify a document.
%% (ii) please avoid comments such as "For a review'', "For some examples",
%% "and references therein" or move them in the text. In general, please leave only references in the bibliography and move all
%% accessory text in footnotes.
%% (iii) Also, please have only one work for each \bibitem.

\nocite{*}

\bibliographystyle{JHEP}
\bibliography{refs}

\end{document}